\documentclass[a4paper,11pt]{article}
\pdfoutput=1 

\usepackage{jinstpub} 
\usepackage{siunitx}
\usepackage{lineno}
\newcommand{\quotes}[1]{``#1''}
\title{\boldmath Novel ion imaging concept based on time-of-flight measurements with low gain avalanche detectors}

\author[a,b,1]{F. Ulrich-Pur,\note{Corresponding author.}}
\author[b]{T. Bergauer,}
\author[c]{A. Hirtl,}
\author[b]{C. Irmler,}
\author[b]{S. Kaser,}
\author[b]{F. Pitters}
\author[d]{S. Rit}

\affiliation[a]{GSI Helmholtzzentrum für Schwerionenforschung GmbH, \\Planckstraße 1, 64291 Darmstadt,
Germany}
\affiliation[b]{Institute of High Energy Physics, Austrian Academy of Sciences,\\
Nikolsdorfer G. 18, 1050 Vienna, Austria}
\affiliation[c]{TU Wien, Atominstitut, \\Stadionallee 2, 1020 Wien, Austria}
\affiliation[d]{Université de Lyon, CREATIS, CNRS UMR5220, Inserm U1294, INSA-Lyon, Université Lyon 1,
Centre Léon Bérard, 69008 Lyon, France}
\emailAdd{F.Ulrich-Pur@gsi.de}
\abstract{Treatment planning in ion beam therapy requires accurate knowledge of the relative stopping power (RSP) distribution within the patient. Currently, RSP maps are obtained via conventional x-ray computed tomography (CT) by converting the measured attenuation coefficients of photons into RSP values for ions. Alternatively, to avoid conversion errors that are inherent to this method, ion computed tomography (iCT) can be used since it allows determining the RSP directly. In typical iCT systems, which usually consist of a tracking system and a separate residual energy detector, the RSP is obtained by measuring the particle trajectory and the corresponding water equivalent path length (WEPL) of single ions travelling through the patient. \\Within this work, we explore a novel iCT approach which does not require a residual energy detector. Instead, the WEPL is estimated indirectly by determining the change in time of flight (TOF) due to the energy loss along the ion's path. For this purpose, we have created a Geant4 model of a TOF-iCT system based on low gain avalanche detectors (LGADs), which are fast 4D-tracking detectors that can measure the time of arrival and position of individual particles with high spatial and time precision. To assess the performance of this TOF-iCT concept, we determined the RSP resolution and accuracy for different system settings using the Catphan\textsuperscript{\textregistered} CTP404 sensitometry phantom. Within the set of investigated system parameters, the lower limit of the RSP accuracy was found at $\SI{0.91}{\percent}$, demonstrating the proof-of-principle of this novel TOF-iCT concept.\\ The main advantage of using this approach is that it could potentially facilitate clinical integration due to its compact design, which, however, requires experimental verification and an improvement of the current WEPL calibration procedure.}


\proceeding{23$^{\text{rd}}$ International Workshop on Radiation Imaging Detectors\\
  June 26-June 30, 2022\\
  Riva del Garda, Italy}

\begin{document}
\maketitle
\flushbottom
\section{Introduction}\label{sec:int}
The rising need for fast particle detectors that are able to simultaneously measure the interaction position and time of arrival (ToA) of a particle with high timing ($\leq \SI{100}{ps}$) and spatial ($< \SI{100}{\micro m}$) precision (i.e. 4D-tracking) has led to the development of low gain avalanche detectors (LGADs). LGADs are silicon detectors with an intrinsic amplification operated at low gain ($\approx \SI{10}{}-\SI{20}{}$) to optimize the signal-to-noise ratio and therefore the time precision \cite{sadrozinski_4d_2018}. Even though LGADs have been mainly employed at high energy physics experiments \cite{lgadusecaserd50}, their radiation hardness and excellent 4D-tracking capabilities make them also suitable detector candidates for ion computed tomography (iCT) \cite{tofctfel}, which is an imaging modality for the direct determination of the relative stopping power (RSP), i.e. energy loss per unit path length expressed relative to the energy loss in water. In conventional iCT systems \cite{johnsonreview}, the RSP is obtained by measuring the particle trajectory and corresponding water-equivalent path length (WEPL) of individual ions passing through the patient using a tracking system and a separate calorimeter placed downstream of the tracker.\\In this study, we present a novel iCT approach, which does not require a residual calorimeter for the WEPL estimation, but instead uses the increase in time-of-flight (TOF) through the patient due to the energy loss along the ion's path. First, we will describe the theoretical background of this method and define a dedicated calibration procedure to map the increase in TOF to the desired WEPL. Finally, we will test this concept by creating an iCT scan of the Catphan\textsuperscript{\textregistered} CTP404 sensitometry phantom \cite{ctp} via a Geant4 \cite{geant4} simulation of a realistic TOF-iCT scanner based on LGADs. Parts of this study are based on the author's PhD thesis \cite{diss}.
\section{Material and Methods}\label{sec:matmeth}
\subsection{Time of flight of ions in matter}
The TOF of an ion with mass $m_0$ travelling along a path $\Gamma$ with the velocity $v({\mathbf{\vec{x}}(s)})$ at position $\mathbf{\vec{x}}(s)$ can be calculated according to
\begin{equation}
\mathrm{TOF} = \int_0^L \frac{\mathrm{d}s}{v({\mathbf{\vec{x}}(s)})} = \int_0^L \frac{\mathrm{d}s}{c \frac{E_{\mathrm{kin}}({\mathbf{\vec{x}}(s)})}{E_{\mathrm{kin}}({\mathbf{\vec{x}}(s)}) + m_0 c^2} \sqrt{1+2 \frac{m_0 c^2}{E_{\mathrm{kin}}({\mathbf{\vec{x}}(s)})}}},
\label{eq:tofgeneral}
\end{equation}
using $c$ as the speed of light, $L=\int_\Gamma \mathrm{d}s$ as the total path length and 
\begin{equation}
\label{eq:ekinrel}
E_{\mathrm{kin}}\left({\mathbf{\vec{x}}(s)}\right)=m_0 c^2 \left( \gamma({\mathbf{\vec{x}}(s)}) -1 \right) \qquad \text{with} \qquad \gamma({\mathbf{\vec{x}}(s)}) = \frac{1}{\sqrt{1- \frac{v^2({\mathbf{\vec{x}}(s)})}{c^2}}}
\end{equation} 
as the relativistic kinetic energy at position $\mathbf{\vec{x}}(s)$. Equation (\ref{eq:tofgeneral}) implies that the TOF strongly depends on the energy loss along the ion's path and therefore the stopping power (SP) of the traversed material. For materials with significant energy loss, the TOF is increased when compared to the TOF in vacuum, where the energy, thus, the velocity can be assumed to be constant ($v({\mathbf{\vec{x}}(s)})=v({\mathbf{\vec{x}}(s=0)})=\mathrm{const}$). In order to understand whether this material-dependent increase in TOF with respect to $\mathrm{TOF}_\mathrm{vac}$ can be exploited for the WEPL determination, it is important to quantify and analyse this effect.
\subsection{Slowing down power}
Using equation (\ref{eq:tofgeneral}) and assuming a small path length $L=\Delta x$, the difference of the real TOF and $\mathrm{TOF}_\mathrm{vac}$ can be determined according to
\begin{equation}
\label{eq:tofdiff}
\Delta \mathrm{TOF}=\mathrm{TOF}-\mathrm{TOF}_\mathrm{vac}=\int_0^{\Delta x} \frac{\mathrm{d}s}{v({\mathbf{\vec{x}}(s)})}-\frac{\Delta x}{v({\mathbf{\vec{x}}(s=0)})}.
\end{equation}
Similar to the stopping power, one can then define the rate of change per unit path length of $\Delta \mathrm{TOF}$, which, from now on, will be referred to as the \quotes{slowing down power} (SDP)
\begin{equation}
\label{eq:sdpdef}
\mathrm{SDP}(E_{\mathrm{kin}}({\mathbf{\vec{x}}(s)}) \coloneqq \frac{\Delta \mathrm{TOF}}{\Delta x}(E_{\mathrm{kin}}({\mathbf{\vec{x}}(s)}).
\end{equation}
\subsection{Relation between slowing down power and stopping power}
\begin{figure}[h]
\begin{center}
\includegraphics[width=0.35\textwidth]{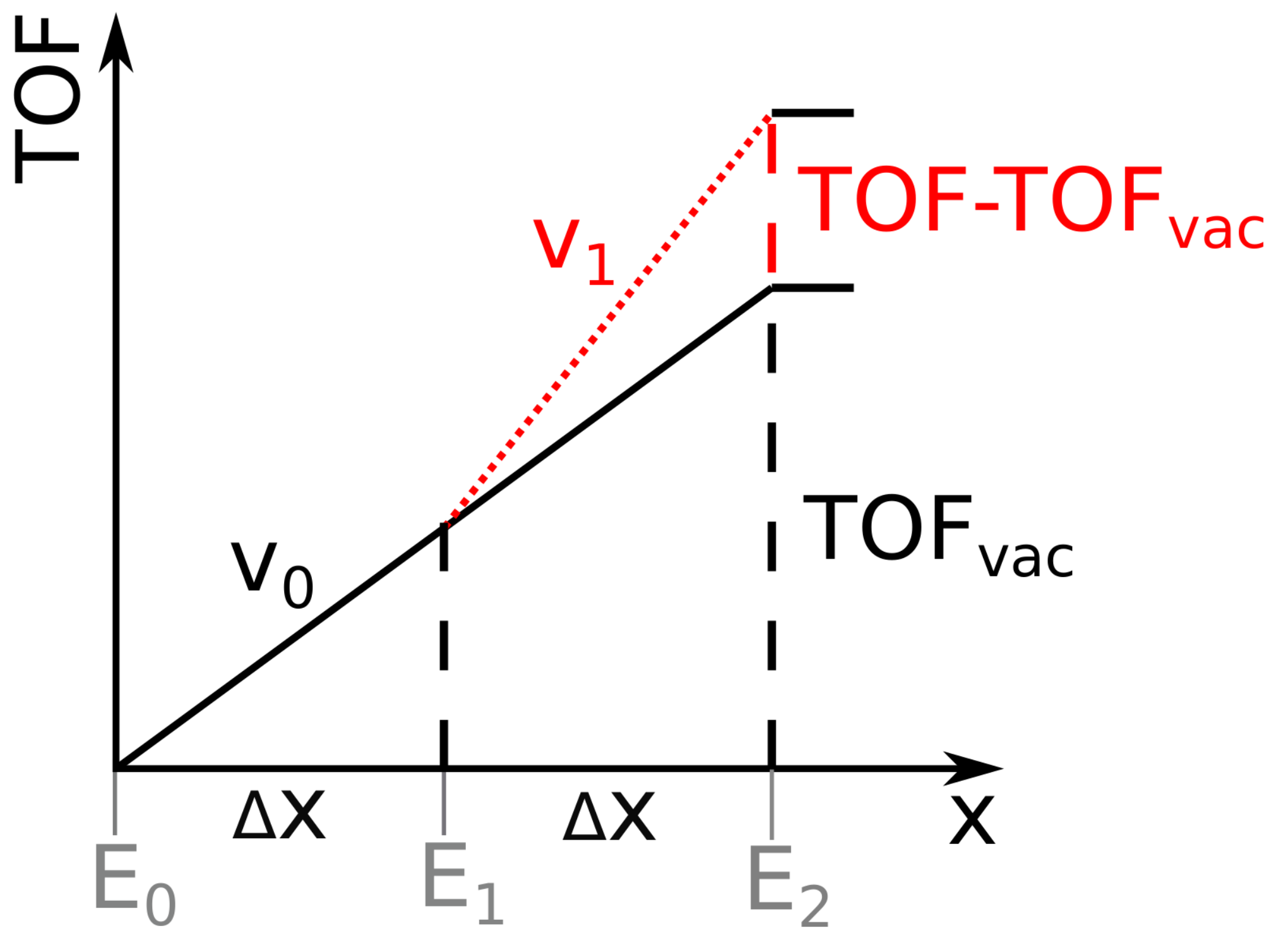}
\end{center}
\vspace{-0.5cm}
\caption[Approximation of the slowing down power.]{Approximation of the slowing down power. The slowing down power corresponds to the increase in TOF (dashed red curve) per unit path length $\Delta x$ caused by the energy loss along the particle's trajectory.}
\label{fig:sdpsketch}
\end{figure} \noindent For an ion with a primary kinetic energy $E_0$ travelling a total path length of $L=2 \Delta x$ in small path length increments $\Delta x$ (figure \ref{fig:sdpsketch}), the true TOF can be approximated by 
\begin{equation}
\label{eq:tofapp}
\mathrm{TOF}=\frac{\Delta x}{v(E(x))}+\frac{\Delta x}{v(E(x+\Delta x))},
\end{equation}
which, after using a first-order Taylor expansion of equation (\ref{eq:tofapp}), reads as
\begin{equation}
\label{eq:tofapp2}
\mathrm{TOF}\approx \frac{2 \Delta x}{v(E(x))} -\frac{ {\Delta x}^2}{v^2(E(x))}\frac{\partial v \left( E(x) \right)}{\partial x}.
\end{equation}
In vacuum, on the other hand, where the energy loss can be neglected ($E(x)=E_0$), equation (\ref{eq:tofapp}) reduces to
\begin{equation}
\label{eq:tofappvac}
\mathrm{TOF}_\mathrm{vac} =\frac{2 \Delta x}{v(E(x))}.
\end{equation}
To estimate the difference between the true TOF and the TOF in vacuum, equation (\ref{eq:tofappvac}) has to be subtracted from equation (\ref{eq:tofapp2}), which results in
\begin{equation}
\label{eq:tofchange}
\mathrm{TOF}-\mathrm{TOF}_\mathrm{vac} \approx -\frac{ {\Delta x}^2}{v^2(E(x))}\frac{\partial v \left( E(x) \right)}{\partial x}.
\end{equation}
To obtain the SDP, equation (\ref{eq:tofchange}) has to be divided by the total path length $L=2 \Delta x$, yielding
\begin{equation}
\label{eq:sdpapp}
\mathrm{SDP}(E(x))=\frac{\mathrm{TOF}-\mathrm{TOF}_\mathrm{vac}}{2 \Delta x} \approx -\frac{ {\Delta x}}{2 v^2(E(x))}\frac{\partial v \left( E(x) \right)}{\partial x}.
\end{equation}
Using the approximation
\begin{equation} 
\frac{\partial v \left( E(x) \right)}{\partial x} = \frac{\partial v \left( E(x) \right)}{\partial E(x)} \frac{\partial E(x) }{\partial x}\approx v'(E(x))\cdot  \mathrm{SP}(E(x))
\end{equation}
with $\frac{\partial E(x) }{\partial x}\approx \mathrm{SP}(E(x))$ as the SP at energy $E(x)$ and $ v'(E(x))=\frac{\partial v \left( E(x) \right)}{\partial E(x)}$ as the partial derivative of equation (\ref{eq:tofgeneral}) with respect to $E(x)$ , the SDP (equation (\ref{eq:sdpapp})) can be rewritten as
\begin{equation}
\label{eq:sdpvssp}
\mathrm{SDP}(E(x))\approx  -\frac{ {\Delta x}}{2 v^2(E(x))} v'(E(x))\cdot \mathrm{SP}(E(x)).
\end{equation}
Similar to the RSP, one can then define the relative slowing down power (RSDP), which is the SDP expressed relative to water
\begin{equation}
\label{eq:rsdp1}
\mathrm{RSDP} \coloneqq \frac{\mathrm{SDP}_{\mathrm{mat}}(E(x))}{\mathrm{SDP}_{\mathrm{H_20}}(E(x))}\approx \frac{-\frac{ {\Delta x}}{2 v^2(E(x))} v'(E(x))\cdot \mathrm{SP}_{\mathrm{mat}}(E(x))}{-\frac{ {\Delta x}}{2 v^2(E(x))} v'(E(x))\cdot \mathrm{SP}_{\mathrm{H_20}}(E(x))}=\mathrm{RSP}.
\end{equation}
Here, the subscripts ``$\mathrm{mat}$'' and  ``$\mathrm{H_20}$'' denote the material for which the corresponding SP and SDP were calculated. Equation (\ref{eq:rsdp1}) indicates, that the $\mathrm{RSP}=\frac{\mathrm{SP}_{\mathrm{mat}}(E(x))}{\mathrm{SP}_{\mathrm{H_20}}(E(x))}$ can be approximated by the RSDP since $v(E(x))$ and $v'(E(x))$ in equation (\ref{eq:rsdp1}) are solely energy-dependent terms, and, therefore, cancel out. To estimate the validity of this approximation, the relative difference between the RSP and RSDP
\begin{equation}
\label{eq:reldiffrsdp}
\epsilon_\mathrm{RSDP}=\frac{\mathrm{RSP}-\mathrm{RSDP}}{\mathrm{RSP}}
\end{equation}
was calculated for different materials and beam energies using the NIST PSTAR database \cite{nistpstar}.
\subsection{Imaging problem and calibration method}
Using equation (\ref{eq:sdpdef}) and equation (\ref{eq:rsdp1}), one can, similar to conventional iCT, define a WEPL
\begin{equation}
\label{eq:WETRSDP}
\mathrm{WEPL}\coloneqq \int_0^{\mathrm{TOF}-\mathrm{TOF}_\mathrm{vac}}\frac{\mathrm{d} \Delta \mathrm{TOF}}{\mathrm{SDP}_{H_20}(\Delta \mathrm{TOF}\left( E(\mathbf{\vec{x}}(s)))\right)} = \int_0^L \mathrm{RSDP}\left( \mathbf{\vec{x}}(s) \right) \mathrm{d}s,
\end{equation}
which can be interpreted as the water equivalent path length resulting in the same total change in TOF as in the target material. Here, the SDP in water was expressed as a function of $\Delta \mathrm{TOF}$ by mapping the total increase in TOF in water for different water thicknesses and a given initial beam energy $E(\mathbf{\vec{x}}(s=0))=E_0$ to the corresponding residual energy $E \left( \mathbf{\vec{x}}(s)\right)$ with $\Delta \mathrm{TOF} \left( E(\mathbf{\vec{x}}(s=0))\right)=0$. This relation between the $\Delta \mathrm{TOF}$ in water and $E \left( \mathbf{\vec{x}}(s)\right)$ can be either obtained by using equation (\ref{eq:tofdiff}) or via a Monte Carlo simulation. After inserting the relation between the RSDP and RSP (equation (\ref{eq:rsdp1})) into equation (\ref{eq:WETRSDP}), it becomes evident that the WEPL as defined in equation (\ref{eq:WETRSDP})   is approximately equal to the conventional definition of the WEPL, which is given by 
\begin{equation}
\mathrm{WEPL}=\int_0^L \mathrm{RSP}\left( \mathbf{\vec{x}}(s) \right) \approx \int_0^L \mathrm{RSDP}\left( \mathbf{\vec{x}}(s) \right) \mathrm{d}s.   
\end{equation}
Therefore, by estimating the particle path and determining the corresponding total change in TOF (equation (\ref{eq:tofdiff})),  the RSP of the scanned object can be obtained using the standard iCT reconstruction algorithms \cite{ddb}. However, since actually measuring the total change in TOF would require precise knowledge of the particle's velocity at each position (equation (\ref{eq:tofdiff})), we first introduce a simpler approach, which approximates equation (\ref{eq:tofdiff}) by the total change in TOF with respect to the TOF in air ($\mathrm{TOF}_\mathrm{air}$), i.e. the measured TOF in air for the same flight distance, but without the target material ($\mathrm{TOF}-\mathrm{TOF}_\mathrm{vac} \approx \mathrm{TOF}-\mathrm{TOF}_\mathrm{air}$). Thus, determining $\Delta \mathrm{TOF}$ reduces to repeating the TOF measurement for a given setup geometry and beam energy after removing the object to be imaged. However, in order to account for the systematic error, which is naturally introduced in this approximation, it is necessary to create a calibration procedure, which maps the measured $\Delta \mathrm{TOF}$ with respect to $\mathrm{TOF}_\mathrm{air}$ to the desired WEPL. Within this study, we have used a fifth-order polynomial as a function of the WEPL to obtain the corresponding calibration curve
\begin{equation}
    \label{eq:calcurve}
    \mathrm{TOF}-\mathrm{TOF}_\mathrm{air} (\mathrm{WEPL,E_0)}\approx \sum_{i=0}^5 a_i (E_0) \cdot \mathrm{WEPL}^i
\end{equation}
with $a_i (E_0)$ as the fit parameters for a given primary beam energy $E_0$. 
\subsection{Experimental setup}\label{sec:exset}
To study the feasibility of this ``sandwich'' TOF-iCT approach, we created a Monte Carlo model of a realistic TOF-iCT scanner based on LGADs using Geant4 version 10.05.1 with the \textit{QGSP\_BIC\_EMY} physics list and \textit{EM\_Options 3}. 
\begin{figure}[htbp]
\centering 
\includegraphics[width=.75\textwidth,origin=c]{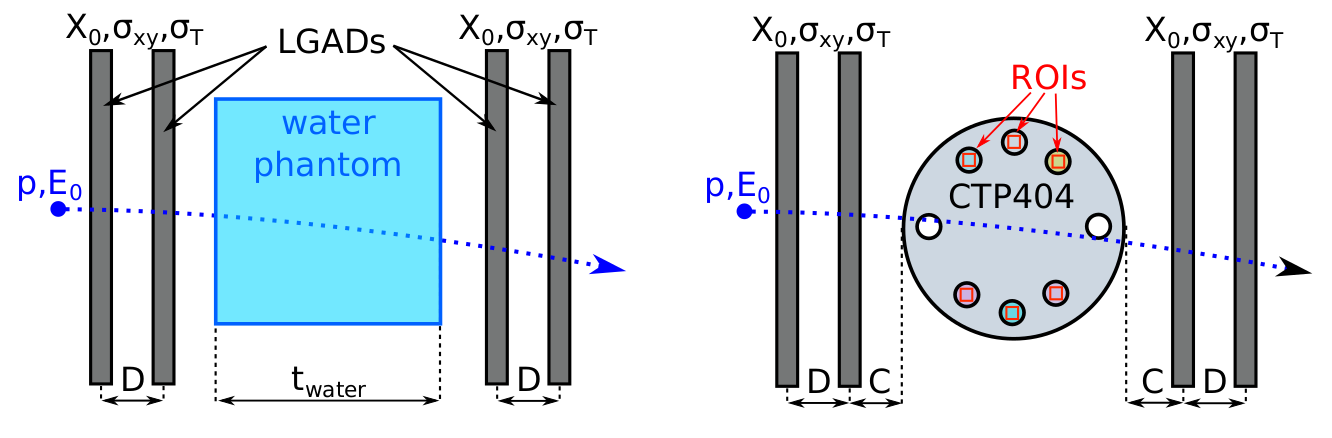}
\vspace{-0.7cm}
\caption{\label{fig:setupsketch} Experimental setup for the calibration (left) and iCT measurement (right).}
\end{figure} \noindent As shown in figure \ref{fig:setupsketch}, the scanner consists of four silicon layers with a thickness of $\SI{300}{\micro m}$ each, simulating generic LGAD detectors. The distance between the individual planes (marked as $D$) and the clearance between the phantom and the innermost planes (marked as $C$) were fixed and set to $\SI{10}{cm}$. To model the intrinsic time and spatial resolution, the ToA and hit position on each sensor were distorted using Gaussian distributions centred at zero and with standard deviations $\sigma_\mathrm{T}$ and $\sigma_{xy}$ corresponding to the time and spatial resolution, respectively. While $\sigma_\mathrm{T}$ was varied between $\SI{0}{}$ and $\SI{100}{ps}$, $\sigma_{xy}$ was set to a fixed value of $\SI[parse-numbers=false]{100/ \sqrt{12}}{\micro m}$, corresponding to a typical sensor pitch of $\SI{100}{\micro m}$. In order to obtain the TOF inside the scanner, the mean ToA in the first two LGAD planes was subtracted from the mean ToA inside the last two LGAD planes. Both the calibration and subsequent iCT scan were performed for the previously mentioned time resolutions and three different proton beam energies, namely $\SI{170}{}$, $\SI{200}{}$ and $\SI{225}{MeV}$. \\For the calibration, we measured the increase in TOF with respect to air (equation (\ref{eq:calcurve})) for different cubic water phantoms with thicknesses ranging from $\SI{0}{}$ to $\SI{200}{mm}$ centred between the two innermost sensors (figure \ref{fig:setupsketch} left). After calculating the calibration parameters according to equation (\ref{eq:calcurve}), iCT scans of the CTP404 phantom (figure \ref{fig:setupsketch} right) were obtained for all investigated beam energies and time resolutions. As shown in figure \ref{fig:setupsketch}, the CTP404 is a PMMA cylinder with a diameter of $\SI{15}{cm}$ featuring six cylindrical inserts with a diameter of $\SI{12.5}{mm}$ each. The materials of the inserts include polymethylpentene (PMP), Teflon, polyoxymethylene (POM, also known as Delrin), Polystyrene, polyethylene (LDPE) and Acrylic. Similar to \cite{tofctfel}, the RSP images were reconstructed via filtered back projection \cite{ddb} using an image voxel size of $\SI{1}{mm^3}$ and 360 projections in $1^\circ$ steps with $\SI{7.5e5}{}$ primary protons each. Then, the RSP values were measured inside each of the six inserts using square-shaped regions of interests (ROIs) with a side length of $\SI{6}{mm}$ (indicated as red boxes in figure \ref{fig:setupsketch}). To estimate the accuracy of the reconstructed RSP, the mean absolute percentage error (MAPE) of the RSP was calculated according to 
\begin{equation}
    \mathrm{MAPE}=\sum_{i=1}^{n=6} \left(\frac{|\mathrm{RSP}_{\mathrm{ref},i}-\mathrm{RSP}_{\mathrm{simu},i}|}{\mathrm{RSP}_{\mathrm{ref},i}}\right) \text{\hspace{-0.25cm}}\left. \frac{}{} \middle/ n \right. ,
\end{equation}
with $\mathrm{RSP}_{\mathrm{simu},i}$ as the mean RSP value of insert $i$ and $\mathrm{RSP}_{\mathrm{ref},i}$ as the corresponding reference RSP value taken from \cite{tofctfel}. As a measure of RSP resolution, the quartile coefficient of dispersion (QCOD) was determined for each insert using
\begin{equation}
    \mathrm{QCOD_\mathrm{RSP}}=\frac{\mathrm{Q}_{3,\mathrm{RSP}}-\mathrm{Q}_{1,\mathrm{RSP}}}{\mathrm{Q}_{3,\mathrm{RSP}}+\mathrm{Q}_{1,\mathrm{RSP}}},
\end{equation}
with $\mathrm{Q}_{1,\mathrm{RSP}}$ and $\mathrm{Q}_{3,\mathrm{RSP}}$ corresponding to the first and third quartiles of the obtained RSP distribution.
\section{Results}\label{sec:res}
\subsection{RSDP error}
\begin{figure}[t]
\centering 
\includegraphics[width=.38\textwidth,origin=c]{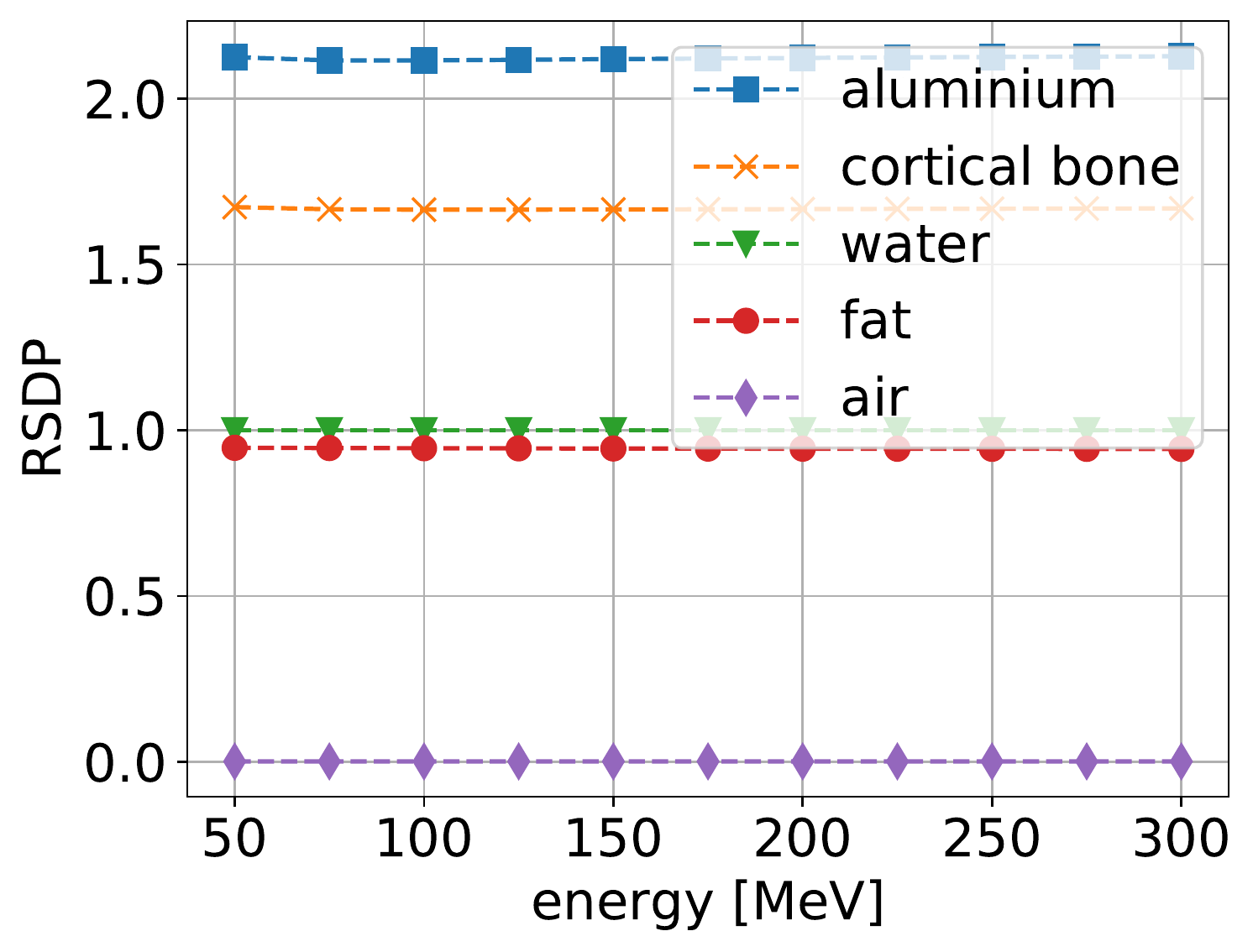}
\qquad
\includegraphics[width=.38\textwidth,origin=c]{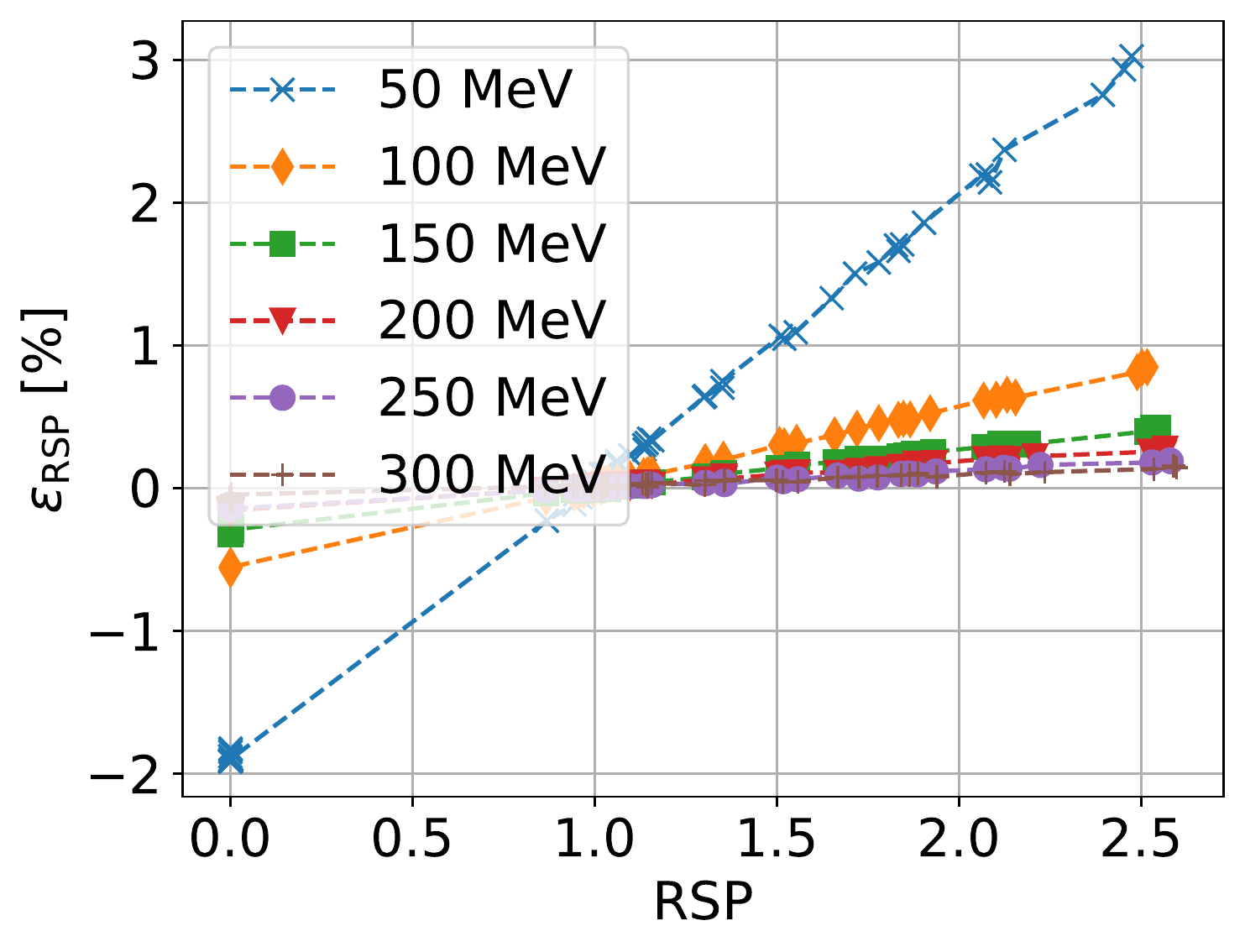}
\vspace{-0.5cm}
\caption{\label{fig:rspdef} RSDP for different materials and beam energies (left) and RSDP error (right).}
\end{figure}

Figure \ref{fig:rspdef} (left) shows the calculated RSDP for different beam energies and materials using the NIST PSTAR database. No significant energy dependence could be observed. Also, the relative difference between the RSDP and RSP, which is depicted on the right, is below $\SI{1}{\percent}$ for beam energies greater than $\SI{100}{MeV}$. For lower beam energies, however, $\epsilon_\mathrm{RSDP}$ ranges between $\SI{-2}{\percent}$ and $\SI{3}{\percent}$, which depends on the investigated RSP value.
\subsection{Ion computed tomography}
\begin{figure}[htbp]
   \centering
\includegraphics[width=.32\textwidth,origin=l]{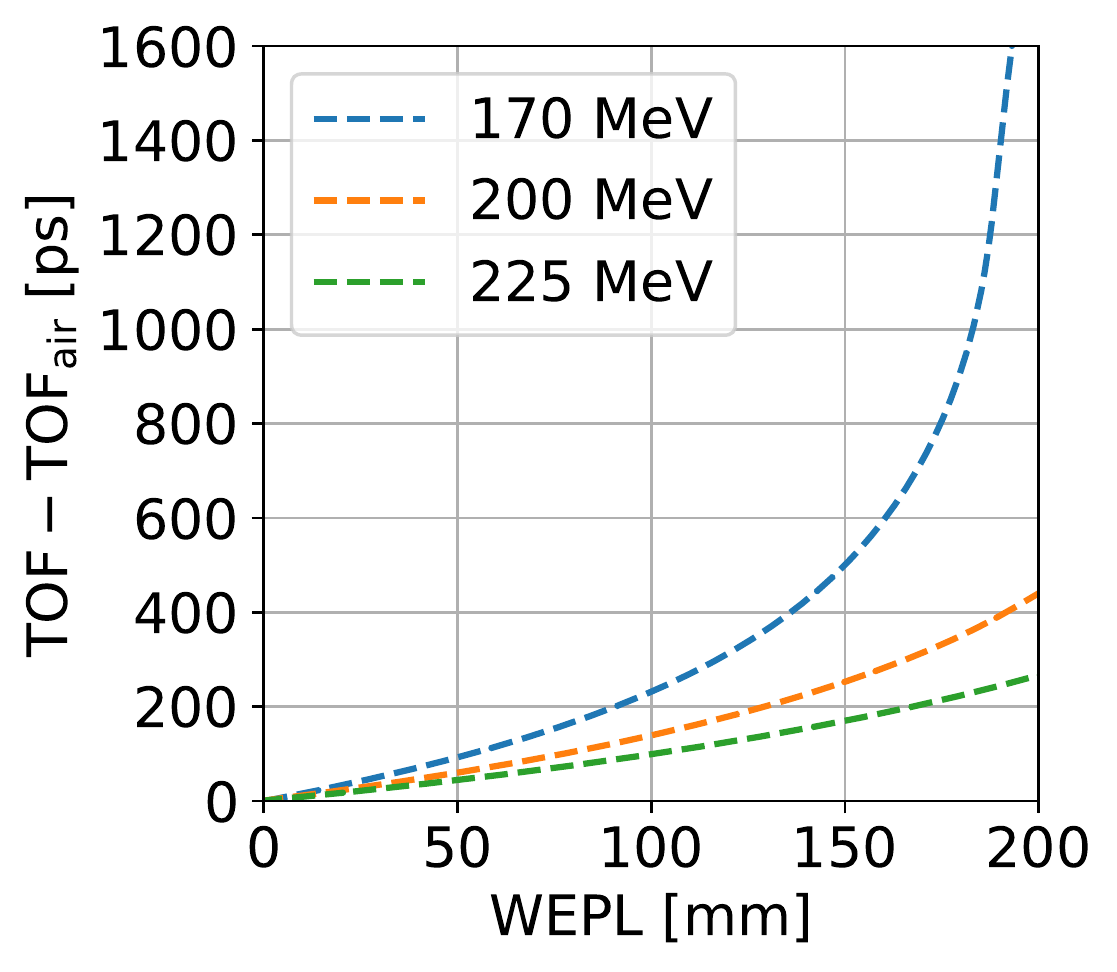}
\qquad
  \includegraphics[width=0.6\textwidth,origin=r]{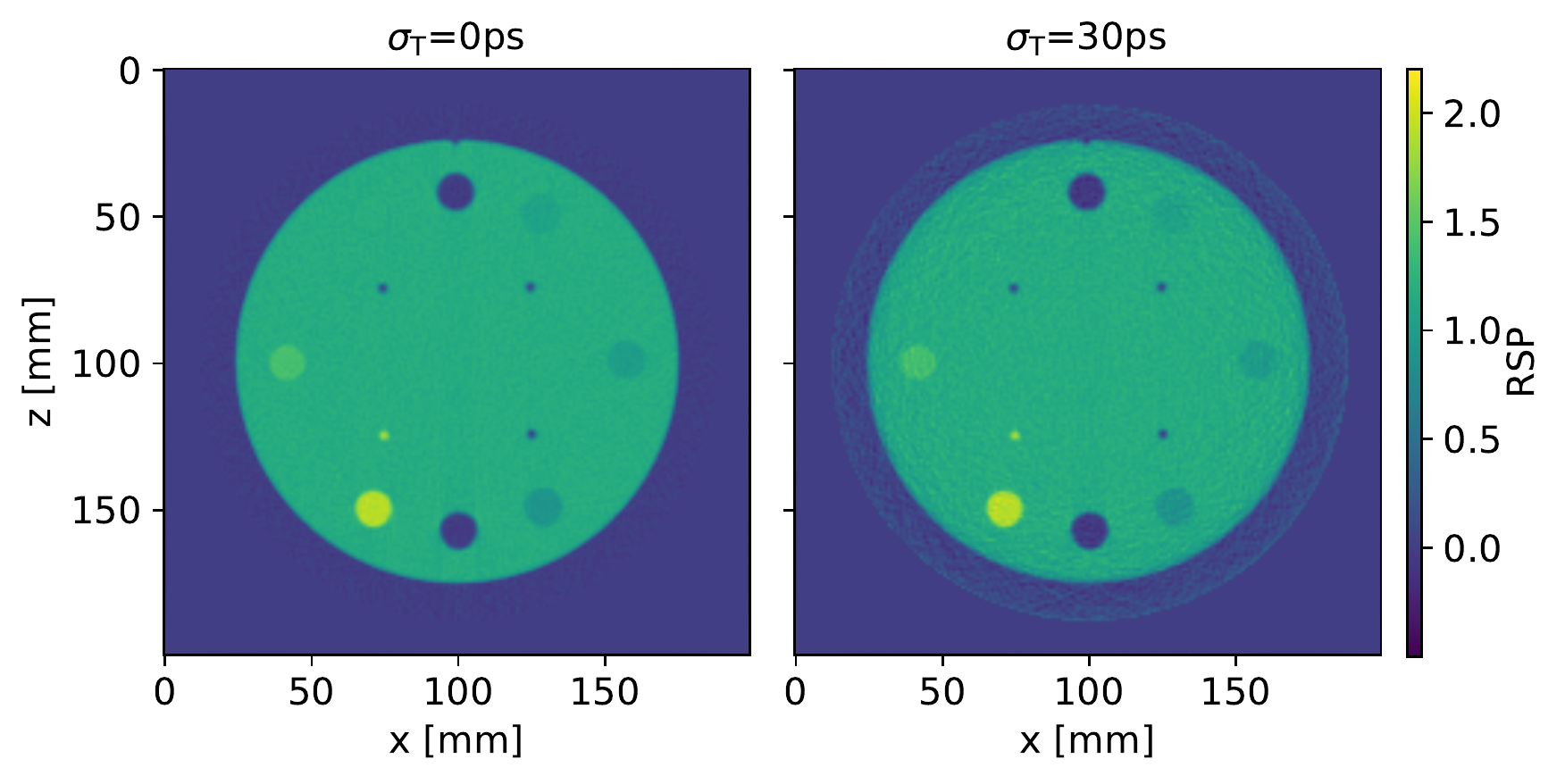}
  \vspace{-0.5cm}
\caption[Calibration curves for $\sigma_\mathrm{T}=\SI{30}{ps}$ and reconstructed central slices of the CTP404 phantom.]{Calibration curves for $\sigma_\mathrm{T}=\SI{30}{ps}$ (left) and reconstructed central slices of the CTP404 phantom recorded with $\SI{175}{MeV}$ protons and $\sigma_\mathrm{T}=\SI{0}{ps}$ (centre) and  $\sigma_\mathrm{T}=\SI{30}{ps}$ (right). The pixels outside the field of view (FOV) have been set to 0.}
\label{fig:allslicessand1}
\end{figure}

\begin{figure}[htbp]
\centering 
\includegraphics[width=.37\textwidth,origin=c]{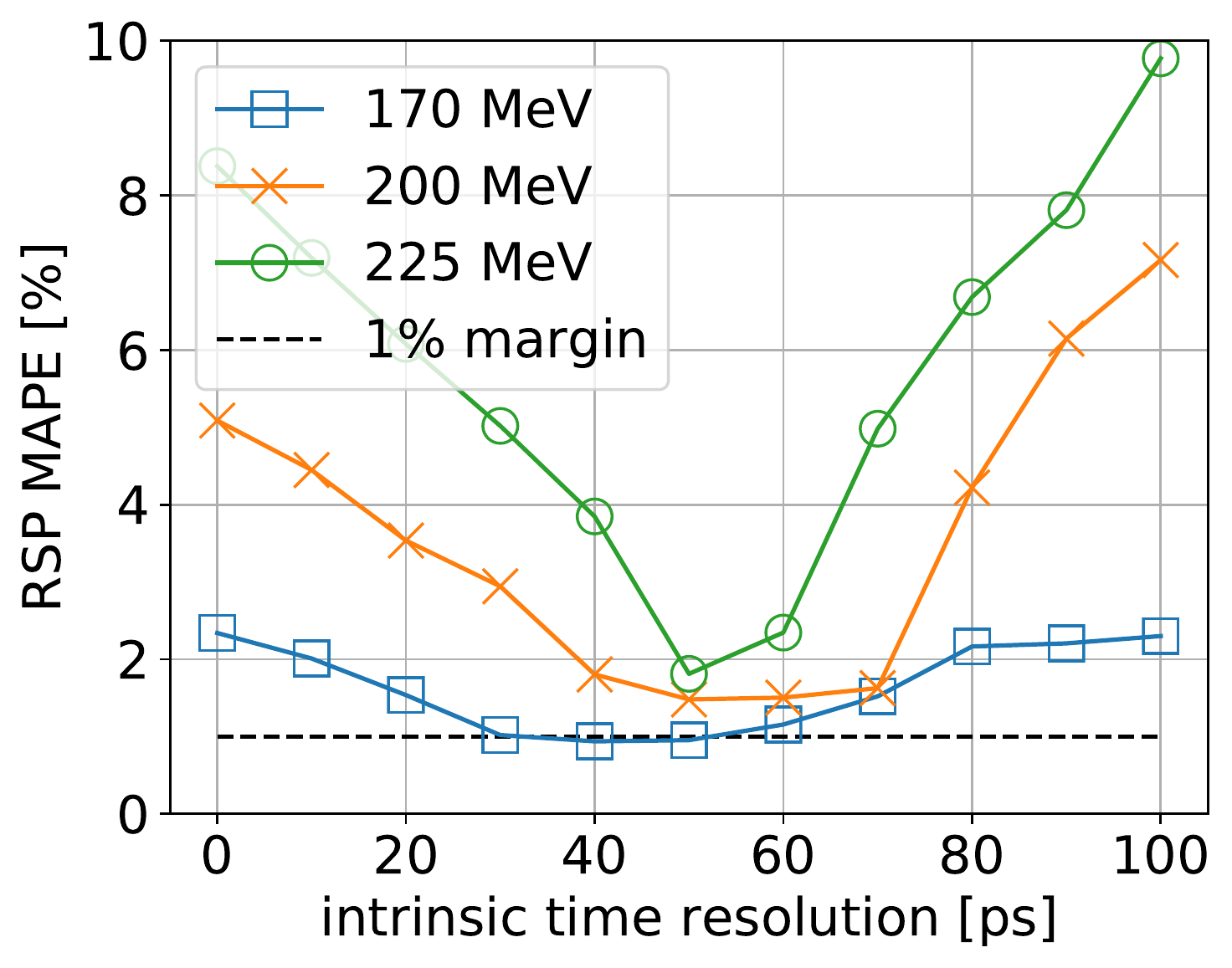}
\qquad
\includegraphics[width=.37\textwidth,origin=c]{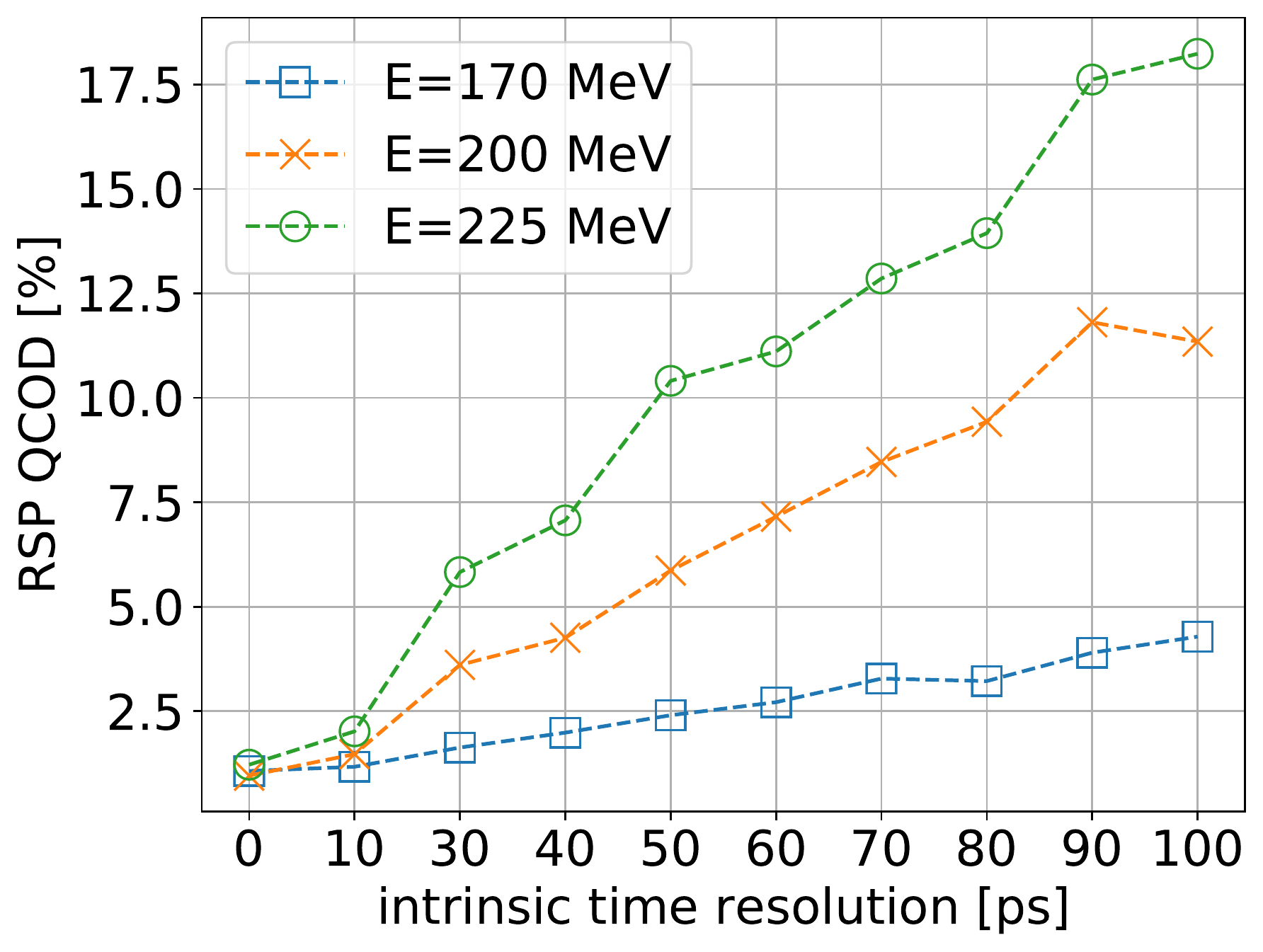}
  \vspace{-0.5cm}
\caption{\label{fig:rspres} RSP MAPE (left) and RSP resolution (right) for different intrinsic time resolutions and primary proton beam energies. The RSP QCOD on the right is only shown for the teflon insert as an example.}
\end{figure}

The absolute TOF difference with respect to air was measured for different water phantom thicknesses as described in section \ref{sec:exset}. The corresponding calibration curves are shown in figure \ref{fig:allslicessand1} (left) for $\sigma_\mathrm{T}=\SI{30}{ps}$.
  \noindent
After determining the calibration parameters as described in section \ref{sec:matmeth}, iCT scans of the CTP404 were created. The corresponding central slices of the iCT images are shown in figure \ref{fig:allslicessand1} (centre and right) for two different system settings. The image noise seems to increase with inferior time resolution. This also becomes apparent when looking at figure \ref{fig:rspres} where the QCOD of the RSP is shown for the Teflon insert (right). As indicated in figure \ref{fig:rspres}, to improve the RSP resolution, the beam energy and $\sigma_\mathrm{T}$ should be kept as low as possible, which is consistent with the calibration curves in Figure \ref{fig:allslicessand1} (left). Also, the RSP accuracy improves with lower beam energy and shows a systematic dependence on the intrinsic time resolution. Only at the lowest of the investigated beam energies and at $\sigma_\mathrm{T}=\SI{50}{ps}$, the RSP MAPE was $\SI{0.91}{\percent}$, which is below the required $\SI{1}{\percent}$ margin \cite{johnsonreview}.
 \noindent
\section{Discussion and Conclusion}\label{sec:disc}
In this study, we could show that ``sandwich'' TOF-iCT could, in principle, be used as an alternative imaging modality to conventional iCT \cite{johnsonreview}. The compact design of this scanner would help to reduce the cost and facilitate the integration into a clinical environment. The performance of this modality, however, was poorer when compared to an iCT systems that uses an LGAD-based TOF calorimeter for the WEPL estimation \cite{tofctfel}, which could be due to the simplified calibration procedure presented in this work. Therefore, more advanced calibration models should be investigated in the future.
\acknowledgments
This project received funding from the Austrian Research Promotion Agency (FFG), grant number 869878.



\end{document}